\documentclass[12pt,preprint]{aastex}
\usepackage{psfig}
\newcommand{\sso}{$^{-1}$}

\slugcomment{To be published in the Astrophysical Journal July 2002}

\begin{document}

\title{Inner Molecular Rings in Barred Galaxies: 
	BIMA SONG CO Observations}
\author{Michael W. Regan\altaffilmark{1,2},
	Kartik Sheth\altaffilmark{2,3,4},
	Peter J, Teuben\altaffilmark{3},
	Stuart N. Vogel\altaffilmark{3},
	}
\altaffiltext{1}{Space Telescope Science Institute, 3700 San Martin Drive,
	Baltimore, MD 21218, mregan@stsci.edu}
\altaffiltext{2}{Visiting Astronomer, Kitt Peak National Observatory,
National Optical Astronomy Observatories, which are operated by the
Association of Universities for Research in Astronomy, Inc. (AURA)
under cooperative agreement with the National Science Foundation}
\altaffiltext{3}{Department of Astronomy, University of Maryland, 
	College Park, MD 20742}
\altaffiltext{4}{Division of Mathematics \& Physical Sciences,
California Institute of Technology,
Pasadena, CA, 91125}

\begin{abstract}

Although inner star-forming rings are common in optical images of
barred spiral galaxies, observational evidence for the accompanying
molecular gas has been scarce. In this paper we present images of
molecular inner rings, traced using the CO (1$-$0) emission line, from
the Berkeley-Illinois-Maryland-Association Survey of Nearby Galaxies
(BIMA SONG).  We detect inner ring CO emission from all five SONG
barred galaxies classified as inner ring (type (r)).  We also examine
the seven SONG barred galaxies classified as inner spiral (type (s)); 
in one of
these, NGC 3627, we find morphological and kinematic evidence for a
molecular inner ring.  Inner ring galaxies have been classified as
such based on optical images, which emphasize recent star formation.
We consider the possibility that there may exist inner rings in which 
star formation efficiency is not enhanced.
However, we find that in NGC 3627 the
inner ring star formation efficiency is enhanced relative to most
other regions in that galaxy.
We note that the SONG
(r) galaxies have a paucity of CO and H$\alpha$ emission interior to
the inner ring (except near the nucleus), while NGC 3627 has
relatively bright bar CO and H$\alpha$ emission; we suggest that
galaxies with inner rings such as NGC 3627 may be misclassified if
there are significant amounts of gas and star formation in the bar.

\end{abstract}

\section{Introduction}

Rings of star formation are a common hallmark of early to intermediate
Hubble type barred spiral galaxies \citep{Buta96}.  The general view
is that gas collects due to gas dissipation in rings located at
various resonances, driven by torques from non-axisymmetric
structures such as bars \citep{Schwarz84,CG85,Byrd94,Buta96}.
Numerous studies indicate that most known rings are sites of
enhanced star formation, implying that star formation is triggered in
these rings \citep{Buta96}.  
Although, without a comparison between the gas in the
rings and the star formation rate, it may be that rings are just
locations with more fuel for star formation.
Consequently, rings are important diagnostic
tools for understanding galaxy dynamics and evolution.

Rings are typically classified into three major
categories \citep{Buta86}: circumnuclear, inner, and outer rings.
Circumnuclear rings are located well interior to the bar, generally
within a kpc of the nucleus.
Inner rings circumscribe the ends of the
stellar bar, and are typically elongated parallel to the bar.  Outer
rings are larger, more diffuse features located well outside the bar
regions, and generally elongated perpendicular to the bar.  Many
studies link circumnuclear, inner, and outer rings to the inner
Lindblad resonance (ILR), ultra-harmonic (also known as 4:1) resonance
(UHR), and outer Lindblad resonance (OLR), respectively \citep{Buta96}.  The
timescale for ring formation varies from 5$\times$10$^9$ years for the
outer rings to only a few $\times$10$^7$ years for the nuclear rings
\citep{Buta96}.
Hence the presence and absence of the different ring structures can
constrain the evolutionary timescale of the galactic disk and its
components, e.g., bars and spiral arms.  Moreover since rings may be
associated with different resonances, the radii of the rings, with
knowledge of the rotation curve, can place constraints on the
pattern speed of the bar.

Circumnuclear rings have prodigious star formation and are most prominent
in blue light \citep{Barth98}; the associated gas is generally not in
a ring morphology \citep{Ken92,Sheth01}.  
Inner rings
are also identified almost exclusively from optical studies tracing recent
star formation, such as blue light excesses or H$\alpha$\ emission.
Similar to circumnuclear rings, inner rings are commonly thought to be
regions where star formation is triggered and enhanced \citep{Buta96}.
However, these conclusions are based on samples in which inner rings
have been identified (i.e. selected) on the basis of excess star
formation in a ring.  It would therefore be useful to investigate
whether there is a population of inner rings which may have remained
undiscovered because the star formation
is not sufficiently remarkable
for the ring to stand out in the optical. Such rings might be
discovered by observing the CO 1$-$0 emission that traces molecular gas.
With an unbiased sample, the question of whether star formation
is enhanced in inner rings can be properly addressed.

Following up on the NRO-OVRO CO survey of the nuclei of galaxies \cite{Sak99},
\cite{Sak00} identified a molecular inner ring in NGC 5005.  However, few
studies to date have had the capability to detect molecular rings.
The problem is that at the distance of most barred galaxies, inner
rings require high (i.e. interferometric) angular resolution to
clearly resolve the inner rings from brighter CO features such as
circumnuclear regions, bar ends, and spiral arms.  
However, the
inner rings lie outside the primary beam of most interferometric
studies, which have generally concentrated on the nuclear region.
Even when larger regions are mosaiced, the spatial dynamic range has
been inadequate to detect inner rings, which, as noted, are fainter
than many other features in barred spirals.  This situation has now
changed with the BIMA CO Survey of Nearby Galaxies (SONG)
\citep{Regan01}.  With BIMA SONG, the disks of 44 nearby spiral
galaxies have been mosaiced with high spatial dynamic range.

\section{Sample Selection and Observations}

As described in \citet{Regan01}, the 44 BIMA SONG sample includes all
galaxies except M33 which have Hubble types between Sa and Sd,
recessional velocity less than 2000 km\,s$^{-1}$, inclination $<
70\degr$, declination $> -20\degr$, and M$_B$ brighter than 11.0.  Of
the 29 SONG galaxies classified as barred (SB or SAB), five have the
(r) classification indicating an inner ring, and seven have the (s)
classification indicating spiral arms and the absence of an inner
ring; the remaining 17 have the (rs) classification indicating a
pseudo-ring or partial ring \citep{VB80,BV83,Buta86}.

The observational techniques and data reduction procedures for the CO
1$-$0 observations of the BIMA SONG galaxies are described in
\cite{Regan01}.  Summarizing, the galaxies were observed with the
10-element Berkeley-Illinois-Maryland Association (BIMA) millimeter
interferometer \citep{Welch}.  Nine of the dozen galaxies discussed in
this paper were observed with a 7-field hexagonal mosaic with a
spacing of 44\arcsec, which gives a half-power field of view of about
190\arcsec.  For NGC 3627, we observed additional fields to cover a
larger area.  NGC 3726 and NGC 4490 were observed with a single
pointing, which yields a half-power field of view of about
100\arcsec.  The angular resolution was typically 6\arcsec.  For NGC
3351 and NGC 3627, we also obtained on-the-fly observations of
CO 1$-$0 with the NRAO 12-m single-dish telescope\footnote{ The National
Radio Astronomy Observatory is operated by the Associated
Universities, Inc., under cooperative agreement with the National
Science Foundation.} and included this data in the maps using a
linear mosaic technique.  
Initial velocity-integrated CO intensity maps 
were derived using the
technique described in \cite{Regan01}; CO velocity files were
obtained using a similar technique.
This technique detected
most of the emission presented. 
To improve the signal to noise ratio we fit rotation curves to these velocities
and generated new velocity-integrated maps by limiting the velocity range in
areas of weak emission to velocities within $\pm$ 50 km s\sso\ of the velocity
predicted by the rotation model.

As part of the SONG study, parallel observations in other wavebands
were obtained.  In this paper we show R band and H$\alpha$ images of
NGC 3344, NGC 3351, NGC 3627, and NGC 3953.  These were obtained at
the 0.9m telescope at Kitt Peak
on the nights of 4-6 April 1999, with
the T2KA 2048x2048 CCD camera in f/13.5 direct imaging mode.  The
integration time for the galaxies in H$\alpha$ ranges from 360 to 420
seconds.  The details of the data reduction for these galaxies are
given in \citet{Sheth01}.  For NGC 3627, we use a K-band image
from \cite{RE97}.

\section{Results}

\subsection{(r) Galaxies}
\begin{figure}
\epsscale{0.45}
\plotone{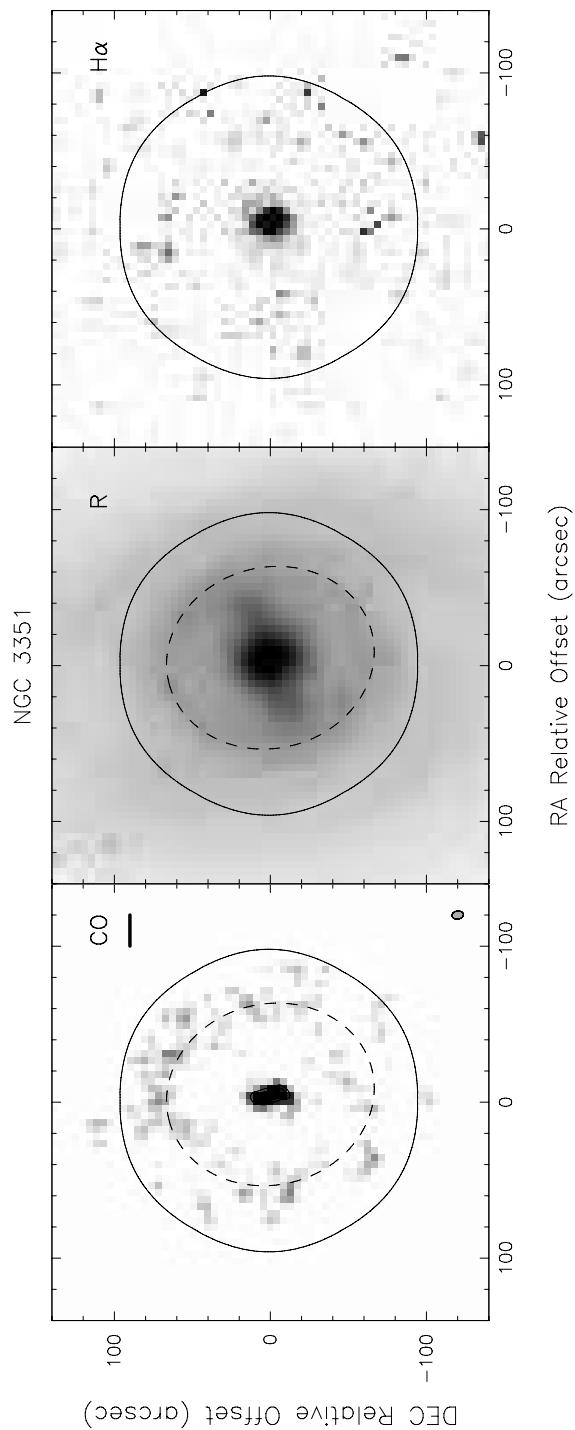}
\caption{ Comparison of CO 1$-$0 (left column), R-band (middle column),
and H$\alpha$ (right column) emission distribution for 
NGC 3344.
The dashed ellipse indicates the location of the inner ring from
\cite{VB80}.
The solid line is the
half-power contour of the BIMA CO mosaic.  The size of the CO
synthesized beam is shown as a filled ellipse in the lower right of
each CO panel.  The thick line in the upper right of each CO panel
indicates a linear distance corresponding to 1 kpc.
}
\label{n3344ring}
\end{figure}

\begin{figure}
\epsscale{0.5}
\plotone{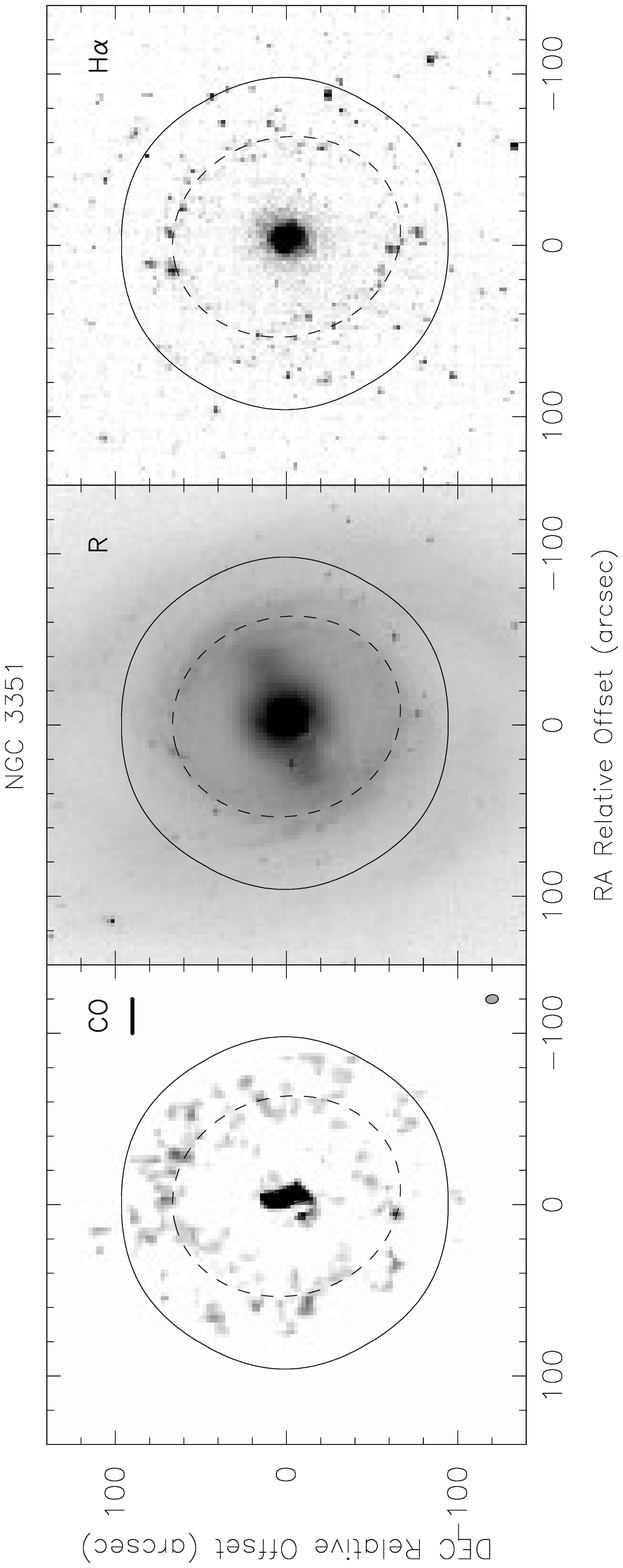}
\caption{ Comparison of CO 1$-$0 (left column), R-band (middle column),
and H$\alpha$ (right column) emission distribution for
NGC 3351. Otherwise the same as Figure \ref{n3344ring}.
}
\label{n3351ring}
\end{figure}

\begin{figure}
\epsscale{0.5}
\plotone{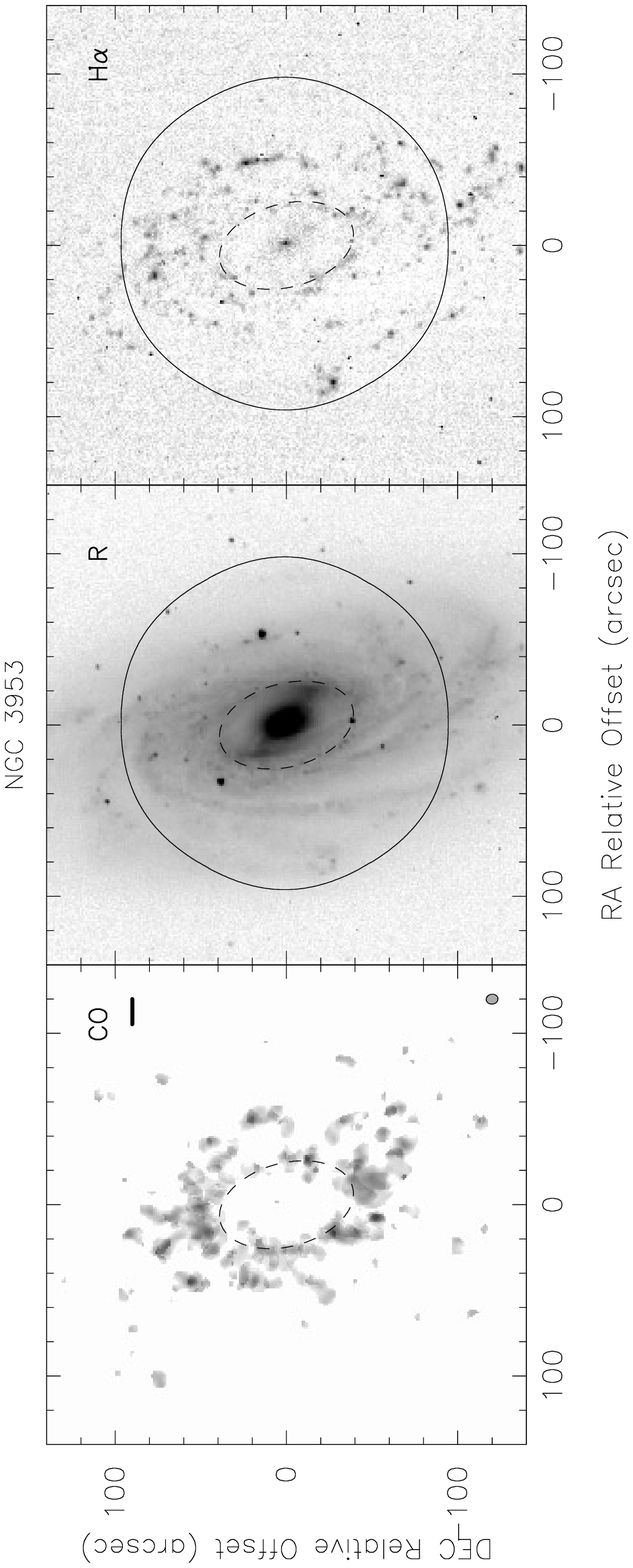}
\caption{ Comparison of CO 1$-$0 (left column), R-band (middle column),
and H$\alpha$ (right column) emission distribution for 
NGC 3953(bottom).  Othewise the same as Figure \ref{n3344ring}.
}
\label{n3953ring}
\end{figure}

Figures \ref{n3344ring}, \ref{n3351ring}, and \ref{n3953ring} show
the BIMA SONG maps of velocity-integrated CO 1$-$0 emission for three of
the galaxies classified as having inner rings 
compared with R band and H$\alpha$ images at the same
scale.  Figures
 \ref{n3344ring}, \ref{n3351ring}, and \ref{n3953ring}
show that CO and H$\alpha$ have similar
distributions.  The solid line indicates the 50\%\ gain contour for
the BIMA CO observations, showing the approximate field of view.  The
dashed line is an ellipse marking the approximate location of the
inner ring.  It is derived using the major axis size determined by de
Vaucouleurs and Buta (1980), and adjusting the ellipse position angle
and ellipticity to approximately match the inner ring seen in the R
band and H$\alpha$ images.  The corresponding ellipse is also plotted
in the CO images.  CO emission coincident with the inner ring is
detected in all three. Although the emission is faint ($\sim$ 1 Jy 
beam\sso\ km s$^{-1}$), it is clearly detected and has the expected galactic
kinematic structure.  In each galaxy, the CO emission also extends
beyond the ring to larger radii, particularly in NGC 3953.  
The CO morphology alone could be described as a hole in the
inner CO disk; it alone does not look very ring-like.
On the
other hand, only in NGC 3351 is emission detected inside of the ring.
In this galaxy, bright emission near the nucleus and
weaker emission in the inner bar region are detected. The central
H$\alpha$ emission is also brighter in this galaxy.  

Inner ring CO emission is also detected in the other two (r) SONG
galaxies, NGC 4725 and NGC 3726, neither of which are shown here
(see \cite{H02}).  For
NGC 4725 the major axis
diameter of the ring is 260\arcsec, larger than the 190\arcsec\
half-power BIMA mosaic region.  Consequently, the sensitivity
at the expected location of the inner ring is poor;
nonetheless, CO emission is detected coincident with the brighter
optical arcs of the inner ring.  The sensitivity is good within the
bar region, but no CO is detected here, except near the nucleus.  In
NGC 3726, CO emission is detected from parts of the inner ring, but
the relatively high inclination, small semi-minor axis of the inner
ring ($\sim$20\arcsec), and presence of nuclear and bar dust lane
CO emission make it difficult to clearly see the CO inner ring.
Nonetheless, some CO emission coincident with the inner ring is
detected.

The CO emission distribution for the five BIMA SONG galaxies
classified as (r) is summarized in Table \ref{Rings}.  The key finding
is that CO emission coincident with the inner ring is detected in all
five, but emission interior to the inner ring is not detected except
near the nucleus.

\subsection{(s) Galaxies}

CO distributions for the seven SONG galaxies classified as spiral (s),
indicating no inner ring, are summarized in the second part of Table
\ref{Rings}(images for all galaxies are shown in \cite{H02}).  
These include four with detected CO emission, but
no obvious organized distribution: NGC 0925, NGC 2403, NGC
4258, and NGC 4490.  The CO emission in NGC 0925 and NGC 2403 is
weak; we note that these are classified as SABd and SABcd
respectively, and are among the latest types in the SONG sample.  
The
CO emission in NGC 4490 and NGC 4258 also has no clearly organized
distribution but is significantly brighter; NGC 4490 is classified as
a peculiar galaxy, and NGC 4258 is a well known disturbed galaxy \citep{VOM72,Krause90}.

NGC 4535 has a CO emission distribution which at first glance appears
to be an inner ring. Detailed inspection, however, shows that the CO
coincides with the two spiral arms, which are very prominent and
tightly wrapped in the optical.  For NGC 4321, CO emission is detected
essentially everywhere in the 3\arcmin\ field of view, including the
expected location of the inner ring.  However, CO emission at the
inner ring location is not enhanced, and for that reason we denote
inner ring emission with a ``?'' in Table \ref{Rings}.  
\subsubsection{NGC 3627}
\begin{figure}
\epsscale{0.4}
\plotone{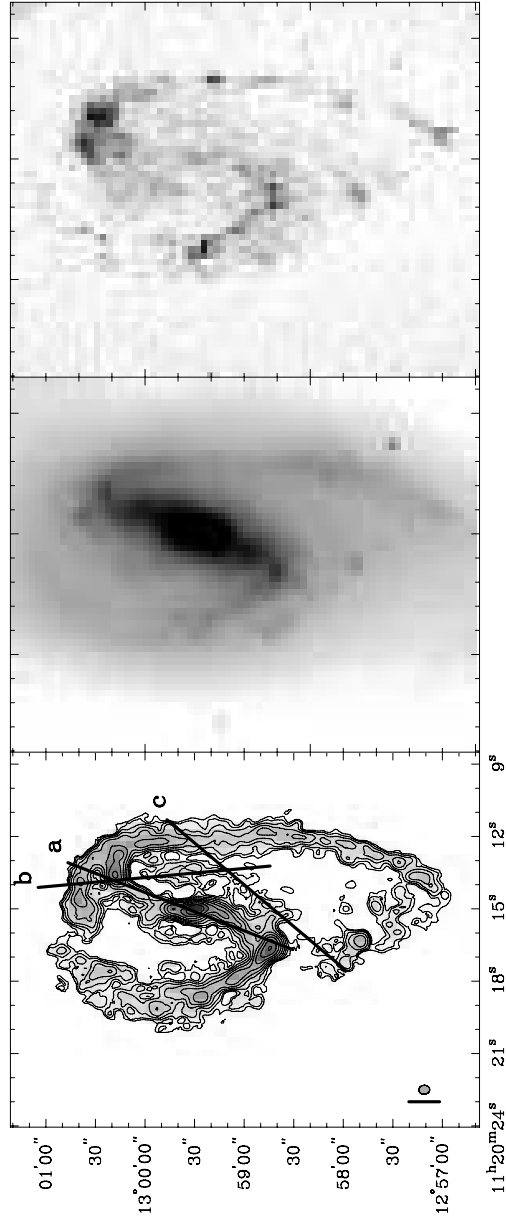}
\caption{Left Panel) Velocity-integrated CO (1$-$0) emission
distribution in NGC
3627, observed as part of the BIMA SONG. Contours begin at 2.5 Jy km
s\sso\ beam$^{-1}$, and are spaced at logarithmic intervals of
1.585.  The vertical bar in the lower left corner shows 1 kpc
at the assumed distance to NGC 3627 (11.1 Mpc).  The synthesized beam
is shown next to the vertical bar.
The three lines show the positions of the long slits used to create
Figure \ref{longslit}.  
Center Panel) A K-band image of NGC 3627 from
\cite{RE97}. The field of view is the same as the CO image.  
Right Panel) H $\alpha$ emission in NGC 3627.}
\label{3627mom}
\end{figure}

The detailed distribution of
velocity-integrated CO emission is shown in Figure \ref{3627mom}.  The
CO emission shows a peak at the nucleus, extends along the leading
edges of the bar, forms two broad peaks at the bar ends, and then
trails off into the spiral arms, with the western arm emission
extending over a greater distance.

The CO map also reveals additional weak emission extended roughly
parallel to the bar but offset approximately 20\arcsec\ both west and
east of the bar.  At its brightest, this emission is more than 10
times weaker than the emission at the bar ends.  Although this
emission is weak, it appears to be resolved, especially in the
brighter western half of the ring.  The deprojected ring is not
obviously elongated.  A deprojected image (e.g. Figure 3.7 in
\citet{Sheth01}) shows that this emission lies along a circle or ring
circumscribing the bar ends.  It is this emission we identify as a
molecular inner ring.

The inner ring is relatively weak compared to the nuclear and bar end
emission.  Such low level emission in the presence of bright emission
in interferometric maps should be viewed with caution due to possible
errors in the deconvolution.  Consequently, we carefully examined the
kinematic structure of the inner ring feature in order to evaluate
whether this structure is in fact an inner ring.

For example, emission that extends over a large velocity range, as it
does in the nucleus of NGC 3627, can appear in the
velocity-integrated maps due to the accumulation of small errors in
individual velocity channels.  These errors leave a distinctive
signature in the data because if the faint emission is due to the
incomplete deconvolution of emission from a bright region, it will have
velocity structure similar to the bright regions.   Therefore, we can
evaluate the weak features by investigating the morphology of the
emission in the velocity cube.

\begin{figure}
\epsscale{0.8}
\plotone{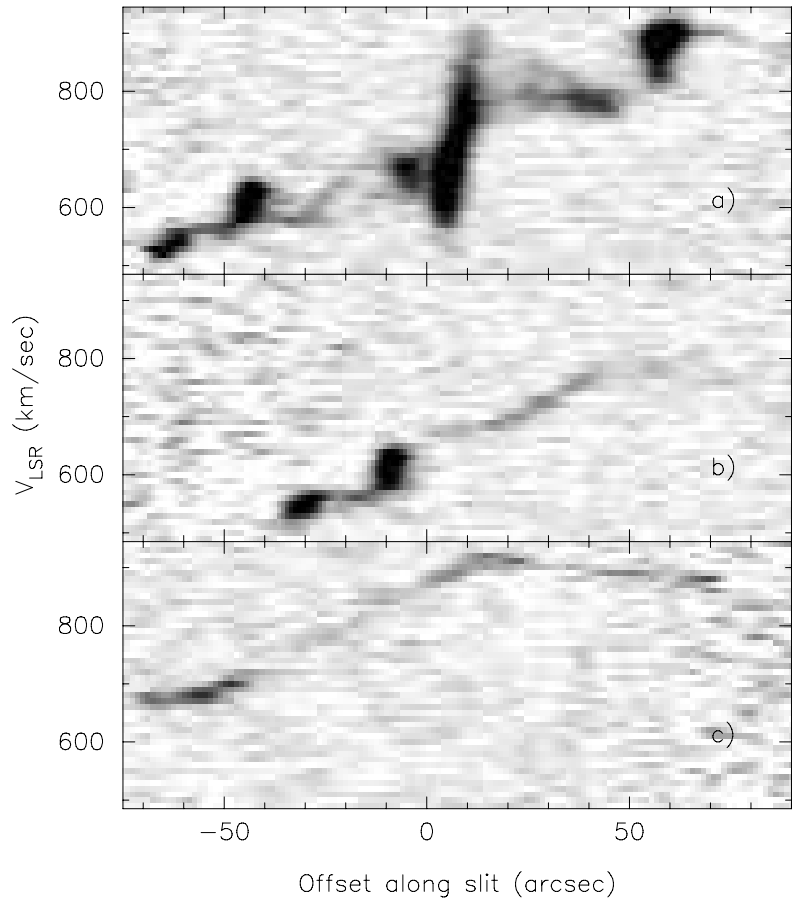}
\caption{
Position-velocity images of CO 1$-$0 emission in NGC 3627.  
Velocity is on the y
axis, position along the ``slit'' on the x axis.  
a) A slit along the bar passing through the bright nuclear
region.  Bright emission from the northern
bar end appear on the left.
b) A slit along the northwestern section of the inner ring.
The ring is at +5\arcsec (North) to +35\arcsec (South).
c) A slit along the southwestern section of the inner ring.
The ring is at -25\arcsec (North) to 0\arcsec (South).
(See Figure \ref{3627mom} for the location of the slits).
Comparing the inner ring emission with the bright emission,
we see that the bar end is at a different velocity, and the nuclear
region emission extends over a much broader velocity range.
}
\label{longslit}
\end{figure}

In Figure \ref{longslit}(b) we show a long slit along the 
northwest section of the ring. 
Emission in the ring is from +5\arcsec\ to +35\arcsec.
Similarly, we show a long slit along the southwest section of
the ring in \ref{longslit}(c).
Here emission in the ring is from -25\arcsec\ to 0\arcsec.
A comparison long slit along the bar
and through the nucleus is shown in \ref{longslit}(a).
The broad, bright emission from the nucleus is clear in the upper panel 
as is the bright emission from the bar ends.
Clearly the ring emission is
very narrow in its velocity extent and quite distinct
kinematicly from both the nucleus and the bar ends.

\begin{figure}
\epsscale{0.9}
\plotone{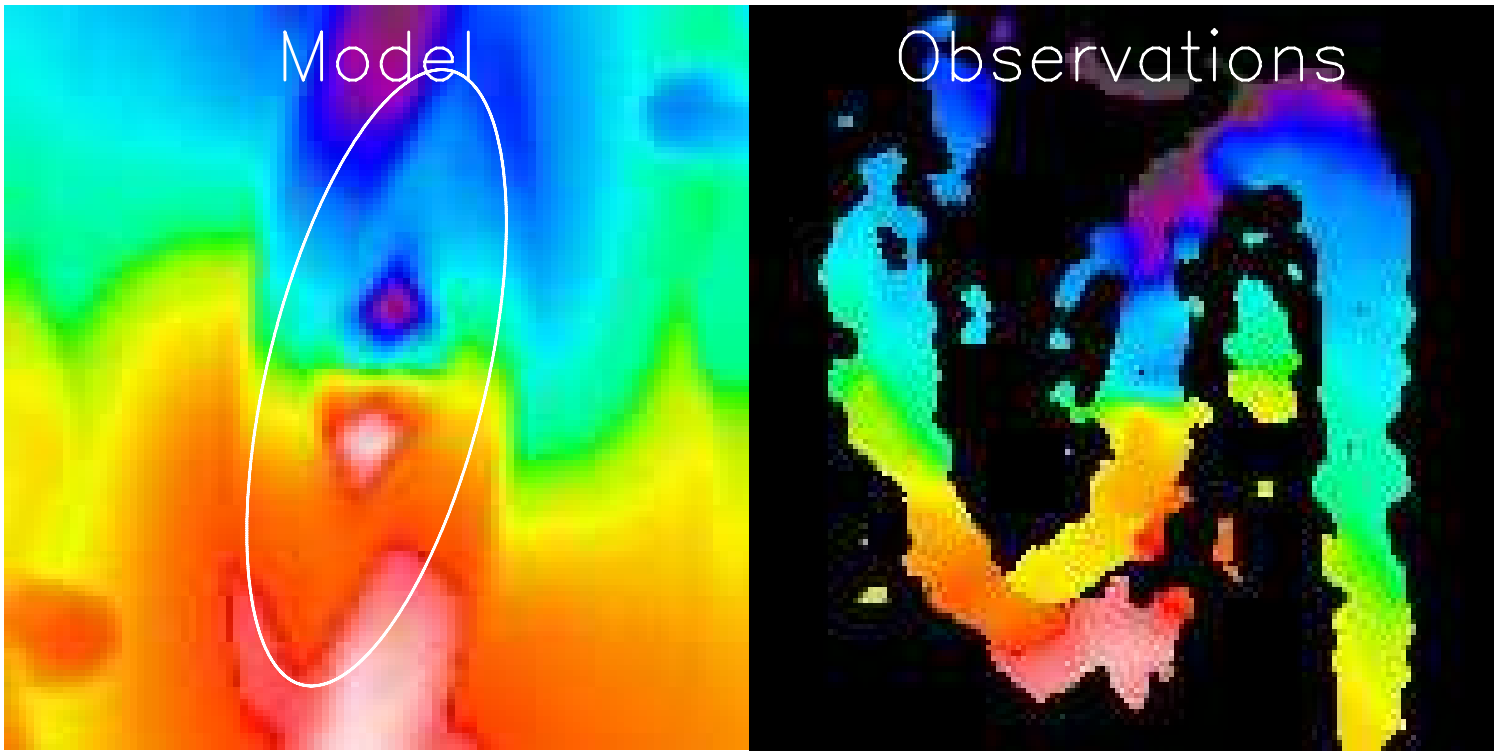}

\caption{
(Right panel) Velocities of CO emission observed for NGC
3627, derived from Gaussian fits to the line profiles. 
(Left panel) Velocities derived from the
standard hydrodynamical model of \cite{PST95}.
 The ellipse
indicates the location of the inner ring.
}
\label{3627vel}
\end{figure}

At this point we have shown that NGC 3627 exhibits CO emission
resembling an inner ring, and the kinematic distribution indicates the
emission is not an artifact.  A further test as to whether the
observed emission is in fact an inner ring is to compare the
kinematics with that predicted by models of barred galaxies.
Hydrodynamic models of gas flow in barred spirals have been successful
at reproducing the kinematic signatures of gas flow in barred
potentials \citep{RVT97,RSV99}.  In Figure \ref{3627vel} (right
panel), we show a velocity map of NGC 3627 made by fitting Gaussian
line profiles to the emission in the cube.  We compare that to one
generated from the models of \cite{PST95} (left panel), where the
model has been smoothed to the resolution of these observations,
rotated, and scaled to match the orientation of NGC 3627 on the sky.
The model is not expected to be an exact match to the observations
because we used one of the standard \citet{PST95} models and did not
attempt to match the gravitational potential of the observed galaxies.
Also, the model does not include star formation, and thus the amount
of gas at various positions will not match the observations (see
discussion in \citet{Sheth00}).  The salient feature of the model is
that, due to the orientations of the bar and the galaxy, gas in the
western half of the inner ring is more red-shifted than gas directly to
the east in the bar (except for the southern nuclear ring gas) or gas
directly to the west further out in the disk; the observed data shows
similarly that the western inner ring gas is more red-shifted than gas
to the east or west. Emission from the
western inner ring is weaker and its kinematics more difficult to
discern. 
Note that these kinematics are also consistent with the model of
the inner ring presented in \cite{Sak00}.
In their model the inner ring was located at the approximate position of
the UHR resonance. 

Summarizing, we have found that although NGC 3627 is classified as an
(s) galaxy, it has a ring of CO emission at the location expected for
an inner ring, with kinematics consistent with that expected for an
inner ring.  

\section{Star Formation and Inner Rings}

\begin{figure}
\epsscale{0.8}
\plotone{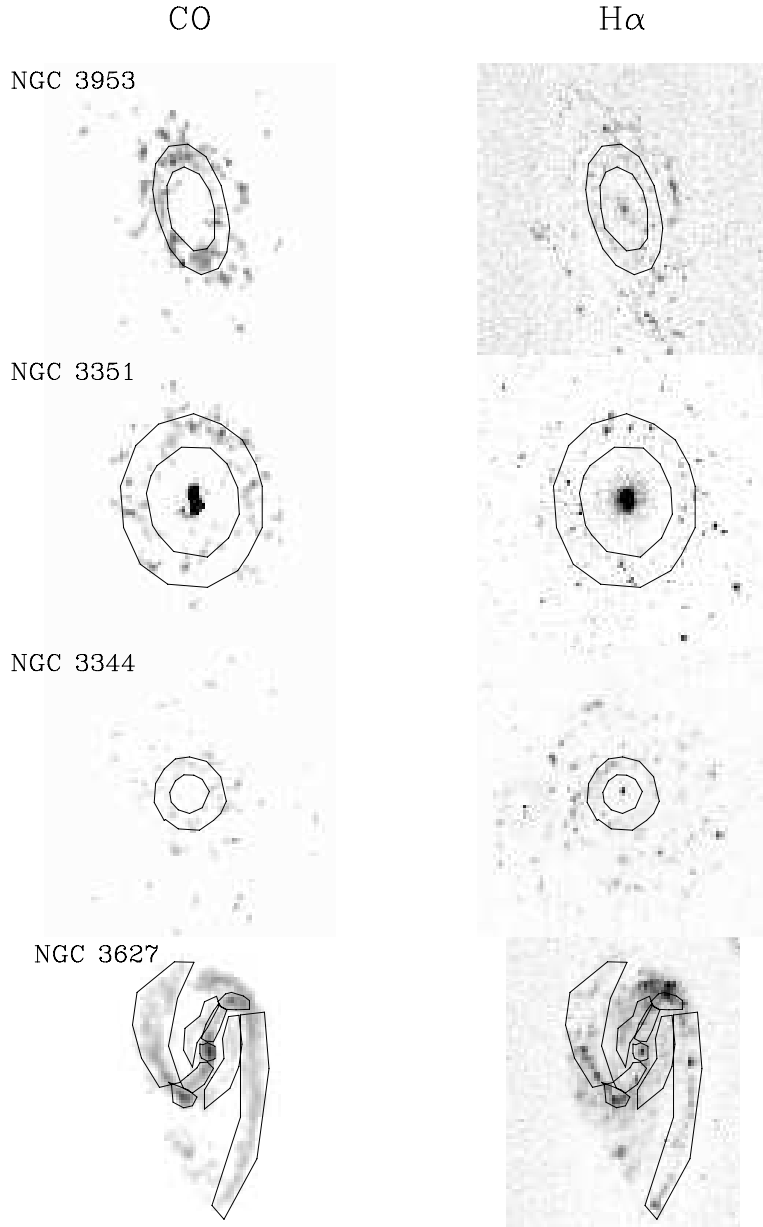}
\caption{
The various apertures used for the photometry. For NGC 3953, NGC 3351, and
NGC 3344 the region used is an annulus.
For NGC 3627 each of the various regions is defined by the drawn polygon.
}
\label{apertures}
\end{figure}

Our detection of a molecular inner ring in a galaxy previously
classified as type (s) suggests that the fraction of galaxies with
inner rings may be underestimated.  As discussed in the Introduction,
inner rings have been historically identified primarily due to star
formation indicators, such as blue optical color excesses.  The
discovery of an inner ring using CO observations
rather than H$\alpha$ might suggest
the existence of a population of inner rings in which the star
formation efficiency (SFE) is lower.  To test this, we can compare the
SFE in the inner ring in the known (r) galaxies with that in NGC 3627.
The SFE can be derived directly from the ratio of the H$\alpha$ surface
brightness to the CO surface brightness.  The three (r) galaxies for
which we have both resolved CO emission in the inner ring and
calibrated H$\alpha$ images are NGC 3344, NGC 3351, and NGC 3953
(shown in Figure 1).  For each galaxy, we defined apertures that
included the inner ring CO and H$\alpha$ emission, and calculated the
average surface brightnesses and ratios listed in Table \ref{ringeff}.
The apertures used for each region are shown in Figure \ref{apertures}
For NGC 3627, because the CO emission is not confined to the inner ring,
we defined separate apertures for the inner ring, bar, bar end, and
spiral arms.  We did not include the bar ends in the region defined as
the inner ring, since here the inner ring would be confused with the
bright spiral arms.  The Western and Eastern inner ring measurements
are listed in Table \ref{ringeff}.  Measurements for other regions
in NGC 3627 are also listed.

Comparing the inner ring measurements listed in Table \ref{ringeff}
we see that the SFE in NGC 3627 is actually the same as that in
the galaxies classified as type (r).  This implies that the optical
classification of NGC 3627 as a type (s) galaxy cannot be attributed
to a paucity of star formation or a lower SFE.  Supporting this, Table
\ref{ringeff} shows that the SFE is higher in the NGC 3627 inner ring
than elsewhere in the galaxy.  The low value in the nucleus might be
attributed to dust extinction of H$\alpha$ emission and enhanced CO
emissivity.   However, the higher SFE of the NGC 3627 ring compared to
the other regions in the galaxy appears significant.
A full investigation of how star formation efficiency varies in different
regions in the BIMA SONG galaxies will be presented in Sheth, Thornley,
Vogel, Regan, Wong (2002, in prep). 

The fact that the inner ring in NGC 3627 does not show up as an excess
of light in the
K-band image may be an indication that the ring is quite young.
If the interaction with NGC 3626 and NGC 3623 has only recently
driven gas into the bar region, then the star formation in the inner
ring may be quite recent and thus not yet had time to create the
older stars that are seen in the near-infrared.

Why was NGC 3627 not identified as an (r) galaxy?  It may be that
the combination of relatively high inclination with the presence of
star formation throughout the bar region makes the inner ring
difficult to discern; in blue light the star formation and dust lanes
within the inner ring would certainly reduce the contrast between the
inner ring and the region interior to it.  
Also, \cite{VB80} showed that the detection rate of inner rings decreased
in high inclination galaxies.
By contrast, in all the (r)
galaxies shown here, in both CO and H$\alpha$ there is a relatively well
defined boundary interior to the radius of the inner ring.  In these
galaxies, the region swept by the bar has very little star formation
(or CO), except near the nucleus, and so the inner ring is easily
detected.  In other words, it may be that inner rings are most
readily {\em detected} in galaxies where the bar has swept most of the gas
from the bar region into the nuclear region.

\section{Summary}

We have detected molecular inner rings in all five of the galaxies in
the BIMA SONG previously classified as having an inner ring.
In addition, we detect an inner ring in NGC 3627, a galaxy previously
thought not to contain an inner ring.
We propose that the large amount of star formation at the bar ends
and along the bar of NGC 3627 combined with the inclination of NGC 3627
reduced the contrast of the star formation in
the inner ring and prevented an optical classification of this galaxy
as having an inner ring.

\acknowledgements

The authors would like to thank the other members of the BIMA SONG
team: Tamara Helfer, Michele Thornley, Tony Wong, Leo Blitz, Douglas
Bock, and Andy Harris for their work on the observations and data
reduction.  
We would also like to thank the comments of the anonymous referee
for comments that improved the paper.
This work is partially supported by NSF
AST-9981289 and by the state of Maryland via support of the Laboratory
for Millimeter-Wave Astronomy.
KS acknowledges support from the NSF grant AST-9981546 which partially
funds the Owens Valley Millimeter Array.

\clearpage	
\begin{deluxetable}{llcccccc}
\tablecolumns{8}
\tablecaption{CO Distribution in (r) and (s) SONG Barred Galaxies\label{Rings}}
\tablehead{
\colhead{Galaxy} & \colhead{Classification} & 
\colhead{Nucleus} & \colhead{Bar} & \colhead{Bar}
& \colhead{Inner} & \colhead{Spiral} & \colhead{Other}\\

& & & \colhead{Dust lanes} & \colhead{Other} & \colhead{Ring} & \colhead{Arms} 

}
\startdata

\cutinhead{Galaxies Classified as Inner Ring (Type (r))}
NGC 3344 & (R)SAB(r)bc       &   &   &   & Y & ?\\
NGC 3351 & SB(r)b;HII        & Y & I &   & Y  & ?\\
NGC 3726 & SAB(r)c           & Y & I & ? & Y  & ?\\  
NGC 3953 & SB(r)bcHII/Lin    &   &   &   & Y  & ?\\
NGC 4725 & SAB(r)ab-pec Sy   & Y &   &   & w & ? \\

\cutinhead{Galaxies Classified as Spiral (Type (s))}

NGC 0925&  SAB(s)d    & &&&&& w\\
NGC 2403&  SAB(s)cd   & &&&&& w\\
NGC 3627&  SAB(s)b Sy & Y & Y & Y & Y & Y \\
NGC 4258&  SAB(s)bc Sy1 & &&&& & Y\\
NGC 4321&  SAB(s)bc HII & Y & Y & Y & ? & Y  \\
NGC 4490&  SAB(s)d pec  & &&&& & Y\\
NGC 4535&  SAB(s)c      & Y & Y & & & Y\\

\enddata
\tablecomments{A ? denotes the inability to determine if a feature exists. A ``w'' means
weak and an ``I'' means inner bar dust lane}
\end{deluxetable}

\clearpage	
\begin{deluxetable}{llcccccc}
\tablecaption{Star Formation Efficiencies in Inner Rings \label{ringeff}}
\tablehead{
\colhead{Galaxy} & \colhead{Region} & \colhead{Area} &
\colhead{H$\alpha$ Flux} & \colhead{CO Flux} &
\colhead{SFR} & \colhead{H$_2$ Mass}& \colhead{SFR/H$_2$}\\
&&kpc${^2}$&erg\,s$^{-1}$\,cm$^{-2}$&Jy\,km\,s$^{-1}$ & 
M$_\sun$\,yr$^{-1}$\,& M$_\sun$ & Gyr$^{-1}$\\
&&&($10^{-13}$) &&& ($10^7$) \\
}
\startdata

NGC 3344 & inner ring &	9.8 & 9.3 & 27 & 0.38 & 4.6 & 8.3 \\
NGC 3351 & inner ring &	25. & 9.9 & 130 & 0.26 & 15 & 1.8 \\
NGC 3953 & inner ring &  19 & 2.3 & 156 & 0.12 & 34 & 0.36 \\
NGC 3627 & E inner ring&	2.6 & 5.2 & 82 & 0.17 & 11  & 1.5 \\
NGC 3627 & W inner ring&	4.6 & 3.5 & 118 & 0.11 & 16  & 0.71 \\
NGC 3627 & nucleus & 	0.7 & 2.4 & 335 & 0.078 & 45 &0.17	\\
NGC 3627 & bar S &	1.6 & 2.6 & 196	& 0.084 &26&0.32\\
NGC 3627 & bar N &	1.5 & 2.5 & 144	&0.082 &19&0.42\\
NGC 3627 & bar end S &	1.2 & 6.9 & 354	&0.23&48&0.47\\
NGC 3627 & bar end N &	1.2 & 8.2 & 304	&0.17&41&0.65\\
NGC 3627 & spiral arm S& 11.9 & 12 &596	&0.42&80&0.52\\
NGC 3627 & spiral arm N& 14.8 &9.1 &691	&0.30&93&0.32\\
\enddata
\tablecomments{The H$_2$ mass was found assuming the standard Galactic
conversion from CO flux to H$_2$ mass.}

\end{deluxetable}

%% ----------------------figureS-----------------------------------------
\clearpage

\clearpage

\begin{thebibliography}{}
\bibitem[Barth et al.(1998)]{Barth98} Barth, A. J., Ho, L. C.,
Filippenko, A. V., Sargent, W. L. W. 1998, \apj, 496, 133
\bibitem[Buta(1986)]{Buta86} Buta, R.\ 1986, \apjs, 61, 609 
\bibitem[Buta \& Combes(1996)]{Buta96} Buta, R., \& Combes, F. 1996,
{\em Fund. Cosmic Physics}, 17, 95.
\bibitem[Buta \& Crocker(1993)]{BC93} Buta, R.\ \& Crocker, 
D.\ A.\ 1993, \aj, 105, 1344 
\bibitem[Buta \& de Vaucouleurs(1983)]{BV83} Buta, R.~\& de 
Vaucouleurs, G.\ 1983, \apjs, 51, 149 
\bibitem[Byrd et al.(1994)]{Byrd94} Byrd, G., Rautiainen, P., 
Salo, H., Buta, R.\ \& Crocker, D.\ A.\ 1994, \aj, 108, 476 
\bibitem[Combes \& Gerin(1985)]{CG85} Combes, F.\ \& Gerin, 
M.\ 1985, \aap, 150, 327 
\bibitem[de Vaucouleurs \& Buta(1980)]{VB80} de Vaucouleurs, 
G.~\& Buta, R.\ 1980, \aj, 85, 637
\bibitem[Helfer et al.(2002)]{H02} Helfer, T. T., Thornley, M. D, 
Regan, M.\ W., Wong, T., Sheth, K., Vogel, S. N., Blitz, L.,
\& Bock, D., C.-J. 2002, \apj,  in prep
\bibitem[Kenney et al.(1992)]{Ken92} Kenney, J.\ D.\ P., 
Wilson, C.\ D., Scoville, N.\ Z., Devereux, N.\ A.\ \& Young, J.\ S.\ 1992, 
\apjl, 395, L79 
\bibitem[Krause et al.(1990)]{Krause90} 
Krause, M., Cox, P., Garcia-Barreto, J.~A., \& Downes, D.\ 1990, \aap, 233, 
L1 
\bibitem[Piner, Stone \& Teuben(1995)]{PST95} Piner, B.\ G., 
Stone, J.\ M.\ \& Teuben, P.\ J.\ 1995, \apj, 449, 508 
\bibitem[Regan \& Elmegreen(1997)]{RE97} Regan, M.\ W.\ \& 
Elmegreen, D.\ M.\ 1997, \aj, 114, 965 
\bibitem[Regan, Sheth \& Vogel(1999)]{RSV99} Regan, M.\ W., 
Sheth, K.\ \& Vogel, S.\ N.\ 1999, \apj, 526, 97 
\bibitem[Regan, Vogel, \& Teuben(1997)]{RVT97} Regan, M.\ W., 
Vogel, S.\ N., \& Teuben, P.\ J.\ 1997, \apjl, 482, L143 
\bibitem[Regan et al.(2001)]{Regan01} Regan, M.\ W., Thornley, M. D.,
Helfer, T. T., Sheth, K., Wong, T., Vogel, S. N., Blitz, L.,
\& Bock, D., C.-J. 2001, \apj,  561, 218
\bibitem[Sakamoto et al.(1999)]{Sak99} Sakamoto, K., Okumura, S.\ K., 
Ishizuki, S.\ \& Scoville, N.\ Z.\ 1999, \apj, 525, 691 
\bibitem[Sakamoto, Baker, \& Scoville(2000)]{Sak00} Sakamoto, 
K., Baker, A.\ J., \& Scoville, N.\ Z.\ 2000, \apj, 533, 149 
\bibitem[Schwarz(1984)]{Schwarz84} Schwarz, M.\ P.\ 1984, 
Proceedings of the Astronomical Society of Australia, 5, 464 
\bibitem[Sheth(2001)]{Sheth01} Sheth, K. 2001, Ph. D. Dissertation, University of Maryland
\bibitem[Sheth et al.(2000)]{Sheth00} Sheth, 
K., Regan, M.\ W., Vogel, S.\ N.\ \& Teuben, P.\ J.\ 2000, \apj, 532, 221 
\bibitem[van der Kruit, Oort, \& Mathewson(1972)]{VOM72} van 
der Kruit, P.~C., Oort, J.~H., \& Mathewson, D.~S.\ 1972, \aap, 21, 169 
\bibitem[Welch et al.(1996)]{Welch} Welch, W.~J.~et al. 1996, \pasp, 108, 93 
\end{thebibliography}
\end{document}